\documentclass{article}
\usepackage{ijcai18}

\usepackage{times}
\usepackage{xcolor}
\usepackage{soul}
\usepackage[utf8]{inputenc}
\usepackage[small]{caption}

\usepackage[hidelinks]{hyperref}

\usepackage{graphicx}
\usepackage{float}

\title{A Virtual Environment with Multi-Robot Navigation, Analytics, and Decision Support for Critical Incident Investigation}

\author{
David L. Smyth,
James Fennell,
Sai Abinesh,
Nazli B. Karimi,\\
\bf{Frank G. Glavin,
Ihsan Ullah,
Brett Drury,
Michael G. Madden}
\\ 
School of Computer Science, National University of Ireland Galway, Ireland\\
michael.madden@nuigalway.ie
}

\begin{document}

\maketitle

\begin{abstract}
Accidents and attacks that involve chemical, biological, radiological/nuclear or explosive (CBRNE) substances are rare, but can be of high consequence. Since the investigation of such events is not anybody's routine work, a range of AI techniques can reduce investigators' cognitive load and support decision-making, including: planning the assessment of the scene; ongoing evaluation and updating of risks; control of autonomous vehicles for collecting images and sensor data; reviewing images/videos for items of interest; identification of anomalies; and retrieval of relevant documentation. Because of the rare and high-risk nature of these events, realistic simulations can support the development and evaluation of AI-based tools. We have developed realistic models of CBRNE scenarios and implemented an initial set of tools.

\end{abstract}

\section{Background and Related Research}
This demonstration presents a new software system that combines a range of artificial intelligence techniques for planning, analysis and decision support for investigation of hazardous crime scenes. It is linked to a new virtual environment model of a critical incident. 
While our work uses a virtual environment as a testbed for AI tools, virtual environments have been more widely used to train responder personnel in near-realistic yet safe conditions. As noted by Chroust and Aumayr \shortcite{chroust2017resilience}, virtual reality can support training by allowing simulations of potential incidents as well as the consequences of various courses of action in a realistic way. Mossel \emph{et al.} \shortcite{mossel2017requirements} provide an analysis of the state of the art in virtual reality training systems which focuses on CBRN disaster preparedness. Use of virtual worlds, such as Second Life and Open Simulator \cite{cohen2013emergency,cohen2013tactical}, have led to improvements in preparation and training for major incident responses. Gautam \emph{et al.} \shortcite{gautam2017human} studied the potential of simulation-based training for emergency response teams in the management of CBRNE victims, and focused on the Human Patient Simulator. \\
\indent CBRNE incident assessment is a critical task which can pose dangers to human life. For this reason, many research projects focus on the use of robots such as Micro Unmanned Aerial Vehicles (MUAV)  to carry out remote sensing in such hazardous environments \cite{marques2017gammaex,baums2017response,daniel2009airshield}. Others include CBRNE mapping for first responders \cite{jasiobedzki2009c2sm} and multi-robot reconnaissance  for detection of threats \cite{schneider2012unmanned}.


\section{Decision Support for Critical Incidents}
This work was undertaken as part of a project called ROCSAFE (Remotely Operated CBRNE Scene Assessment and Forensic Examination), see Drury \emph{et al.} \shortcite{rocsafe}. As described in the following sections, we have implemented a baseline set of decision support systems and developed 3D world models of CBRNE incidents using a physics-based game engine to model Robotic Aerial Vehicles (RAVs). 
\\
\indent The operator can issue a command to have the RAVs survey the scene. The RAVs operate as a multi-agent robot swarm to divide up work between them, and relay information from their sensors and cameras to a central hub. There, our Image Analysis module uses a Deep Neural Network (DNN) to detect and identify relevant objects in images taken by RAV cameras. It also uses a DNN to perform pixel-level semantic annotation of the terrain, to support subsequent route-planning for Robotic Ground-based Vehicles (RGVs). Our Probabilistic Reasoning module assesses the likelihood of different threats, as information arrives from the scene commander, survey images and sensor readings. Our Information Retrieval module ranks documentation, using TF-IDF, by relevance to the incident. All interactions are managed by our purpose-built JSON-based communications protocol, which is also supported by real-world RAVs, cameras and sensor systems. This keeps the system loosely coupled, and will support future testing in real-world environments. 

\subsection{Contributions}
The key contributions of this work are: (1) an integrated system for critical incident decision support, incorporating a range of different AI techniques; (2) an extensible virtual environment model of an incident; and (3) a communications protocol for information exchange between subsystems. These tools are open-source and are being released publicly. This will enable other researchers to integrate their own AI modules (e.g. for image recognition, routing, or probabilistic reasoning) within this overall system, or to model other scenarios and evaluate the existing AI modules on them.

\subsection{3D Model of Criticial Incident}
\begin{figure}[H]
    \centering
    \includegraphics[width=3.3in]{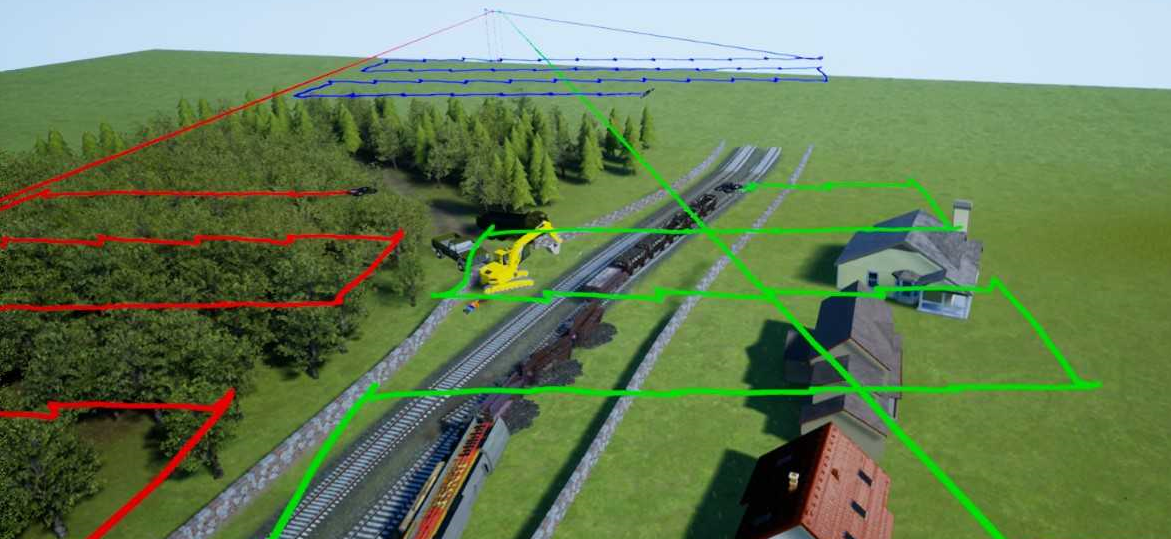}
    \caption{RAV flight paths for surveying and data collection.}
    \label{fig:rav}
\end{figure}

To facilitate the development and testing of the software system, we have designed and publicly released a virtual environment \cite{repo}. It is built with Unreal Engine, a suite of integrated tools for building simulations with photo-realistic visualizations and accurate real-world physics. UE is open source, scalable and supports plugins that allow the integration of RAVs and RGVs into the environment. We chose an operational scenario to model that consists of a train carrying radioactive material in a rural setting. This is shown in Figure \ref{fig:rav}. We used Microsoft's \emph{AirSim} \cite{airsim2017fsr} plugin to model the RAVs. AirSim exposes various APIs to allow fine-grain control of RAVs, RGVs and their associated components. We have replicated a number of APIs from real-world RAV and RGV systems to facilitate the application of our AI tools to real-world critical incident use-cases in the future, after testing in the virtual envionment. 
\subsection{Communications}
We developed a purpose-built JSON-format protocol for all communications between subsystems. There are a relatively small number of messages which are sent at pre-defined intervals, so we use a RESTful API \cite{Richardson:2013:RWA:2566876}.
Our communications protocol is designed to support various vehicles with significant autonomy and is flexible enough to integrate with various components, using different standards, protocols and data types. In this demonstration, we concentrate on RAVs. Since decision making may happen within each RAV’s single-board computer, we have also facilitated direct communication between the RAVs. 
\subsection{Autonomous Surveying and Image Collection}
Our multi-agent system supports autonomous mapping of the virtual environment. This involves discretizing a rectangular region of interest into a set of grid points. At each of the points, the RAV records a number of images and metadata relating to those images. Four bounding GPS coordinates, that form the corner points of a rectangle, can be passed in through a web-based interface. 
\\ \indent Our initial route planning algorithm develops agent routes at a centralized source and distributes the planned routes to each agent in the multi-agent system \cite{gem}. Our current implementation uses a greedy algorithm, which generates subsequent points in each agent’s path by minimizing the distance each agent needs to travel to an unvisited grid point. Current state of the art multi-agent routing algorithms use hyper-heuristics, which out-perform algorithms that use any individual heuristic \cite{hyper}. We intend to integrate this approach with learning algorithms such as Markov Decision Processes \cite{ulmer2017route} in order to optimize the agent routes in a stochastic environment, for example where RAVs can fail and battery usage may not be fully known.
\subsection{Image Processing and Scene Analysis}
Our Central Decision Management (CDM) system uses a DNN to identify relevant objects in images taken by the RAV cameras. Specifically, we have fine-tuned the object detection model \emph{Mask R-CNN} \cite{he2017mask}, using annotated synthetic images that we collected from our virtual scene. This kind of training on a synthetic dataset has been shown to transfer well to real world data: in self-driving cars \cite{pan2017virtual}, pedestrian detection \cite{hattori2015learning}, 2D-3D alignment \cite{aubry2014seeing} and object detection \cite{tian2018training}.
\\ \indent \emph{Mask R-CNN} is an extension of \emph{Faster R-CNN} \cite{NIPS2015_5638} and is currently a state-of-the-art object detection DNN model that not only detects and localizes objects with bounding boxes, but also overlays instance segmentation masks on top, to show the contours of the objects within the boxes. The goal of this is to draw the crime scene investigator's attention to objects of interest within the scene, even if there are overlapping objects, and the object labels are an input into the probabilistic reasoning module. We plan to further enhance the performance of this technique by retraining/fine-tuning the network on other relevant datasets, for example, Object deTection in Aerial (DOTA) images \cite{xia2017dota}.

\subsection{Reasoning and Information Retrieval}
To synthesize data and reason about threats over time, we have developed a probabilistic model in BLOG \cite{blog}. Its goal is to estimate the probabilities of different broad categories of threat (chemical, biological, or radiation/nuclear) and specific threat substances, as this knowledge will affect the way that the scene is assessed. For example, initially, a first responder with a hand-held instrument may detect evidence of radiation in some regions of the scene. Subsequent RAV images may show damaged vegetation in those and other regions, which could be caused by radiation or chemical substances. RAVs may be dispatched with radiation sensors to fly low over those regions, subsequently detecting a source in one region. Using keywords that come from sources such as the object detection module, the probabilistic reasoning module, and the crime scene investigators, the CDM retrieves documentation such as standard operating procedures and guidance documents from a knowledge base. It ranks them in order of relevance to the current situation, using Elastic Search and a previously-defined set of CBRNE synonyms. These documents are re-ranked in real-time as new information becomes available.
\section*{Acknowledgements}
This research is funded by the European Union's Horizon 2020 Programme under grant agreement No. 700264.
\bibliographystyle{named}
\bibliography{ijcai18}





\end{document}